\documentclass[aps,prb,amsmath,amssymb,amsfont,showpacs,superscriptaddress,reprint]{revtex4-1}
\usepackage{amsmath}
\usepackage{bm}
\usepackage[dvipsnames]{xcolor}
\usepackage{braket}
\usepackage[artemisia]{textgreek}  
\usepackage{float}
\usepackage[pdftex]{graphicx}      

\begin{document}

\title{Improved laser based photoluminescence on single-walled carbon nanotubes}

\author{S. Kollarics}
\affiliation{Department of Physics, Budapest University of Technology and Economics and MTA-BME Lend\"{u}let Spintronics Research Group (PROSPIN), P.O. Box 91, H-1521 Budapest, Hungary}

\author{J. Palot\'{a}s}
\affiliation{Department of Physics, Budapest University of Technology and Economics and MTA-BME Lend\"{u}let Spintronics Research Group (PROSPIN), P.O. Box 91, H-1521 Budapest, Hungary}

\author{A. Bojtor}
\affiliation{Department of Physics, Budapest University of Technology and Economics and MTA-BME Lend\"{u}let Spintronics Research Group (PROSPIN), P.O. Box 91, H-1521 Budapest, Hungary}

\author{B. G. M\'{a}rkus}
\affiliation{Department of Physics, Budapest University of Technology and Economics and MTA-BME Lend\"{u}let Spintronics Research Group (PROSPIN), P.O. Box 91, H-1521 Budapest, Hungary}

\author{P. Rohringer}
\affiliation{Faculty of Physics, University of Vienna, Strudlhofgasse 4., Vienna A-1090, Austria}

\author{T. Pichler}
\affiliation{Faculty of Physics, University of Vienna, Strudlhofgasse 4., Vienna A-1090, Austria}

\author{F. Simon}
\affiliation{Department of Physics, Budapest University of Technology and Economics and MTA-BME Lend\"{u}let Spintronics Research Group (PROSPIN), P.O. Box 91, H-1521 Budapest, Hungary}


\begin{abstract}Photoluminescence (PL) has become a common tool to characterize various properties of single-walled carbon nanotube (SWCNT) chirality distribution and the level of tube individualization in SWCNT samples. Most PL studies employ conventional lamp light sources whose spectral distribution is filtered with a monochromator but this results in a still impure spectrum with a low spectral intensity. Tunable dye lasers offer a tunable light source which cover the desired excitation wavelength range with a high spectral intensity, but their operation is often cumbersome. Here, we present the design and properties of an improved dye-laser system which is based on a Q-switch pump laser. The high peak power of the pump provides an essentially threshold-free lasing of the dye laser which substantially improves the operability. It allows operation with laser dyes such as Rhodamin 110 and Pyridin 1, {\color{black}which are otherwise on the border of operation of our laser}. Our system allows to cover the 540-730 nm wavelength range with 4 dyes. In addition, the dye laser output pulses closely follow the properties of the pump this it directly provides a time resolved and tunable laser source. We demonstrate the performance of the system by measuring the photoluminescence map of a HiPco single-walled carbon nanotubes sample.\end{abstract}

\maketitle

\section{Introduction}

Photoluminescence spectroscopy on SWCNTs \cite{OconnellSCI} has proven to be an invaluable tool in characterizing the chirality distribution in samples \cite{Bachilo:Science298:2361:(2002)}, understanding the fundamental nature of light induced excitations \cite{HeinzSCI2005,MaultzschPRB2005}, and to study various phenomena including the role of electron-phonon coupling \cite{PhysRevLett.94.127402}, doping \cite{OconnellPRL2004}, and intertube interactions \cite{DoornJACS}. Photoluminescence measurements require a stable light source with tunable wavelength, which has the sufficient brilliance (i.e. output power per unit wavelength). Most photoluminescence spectrometers operate with arc-discharge or thermal radiation sources that can have brilliance as low as a few mW per nm. Clearly, tunable lasers are appropriate for the purpose of a high brilliant source, however, their power stability is relatively modest and their handling and tuning requires great skill and experience and it is more difficult, albeit not impossible to automatize wavelength selection \cite{duarte1990dye}.

Since the first applications of dye molecules as lasing material \cite{sorokin,schafer1966} dye lasers found their place in industry and research as versatile and reliable light sources. There are more than a hundred dyes which are known for their appropriate lasing properties and commercially available. Using them for the whole spectral range from UV to NIR is possible with appropriate pump sources. However, operation of a dye laser is often difficult in the wavelength range close to the pump wavelength, moreover for some dyes which are known to have a modest quantum yield. In addition, output power noise (or power fluctuation) in pumped lasers is known to originate from the competition between the photons amplified from spontaneous emission versus the stimulated emission \cite{duarte1990dye}. It is expected that a tunable laser pumped by a pulsed laser source with the same average power, might outperform their continuous wave counterparts due to the inherently high peak power during the pulses. Since laser operation is a highly non-linear phenomenon, it is expected that laser operation becomes more robust when pumped by a pulsed light source.

Here, we report the construction and performance characterization of a tunable dye laser system pumped by a Q-switch laser. The system is based on a continuous wave (CW) dye laser which was already available. Our technical development essentially involves replacing the CW pump laser by a Q-switch pump laser. We demonstrate that the high peak power during pulses results in a threshold-free operation for the pump power and it also allows the use of dyes, {which are otherwise on the border of operation of our CW laser}. We find the optimal settings for the repetition rate of the pulsed pump laser that controls the peak power for a given average power. The system allows to perform tunable laser based photoluminescence measurements in the 540-730 nm wavelength range, which is demonstrated by measurements on single-walled carbon nanotubes.

\section{Experimental}

\paragraph{Optical Setup}
The dye laser system is based on an optical resonator (Radiant Dyes Laser \& Acc. GmbH, Germany) pumped by a frequency doubled pulsed (using a Q-switch) Nd:YLF laser (Coherent Evolution 15) with $\lambda=527\,\text{nm}$ and for comparison also by a frequency doubled continuous wave (CW) Nd:YAG laser (Coherent Verdi 5G) with $\lambda=532\,\text{nm}$. We used four different dyes (purchased from Radiant Dyes) to cover a broad wavelength range. The dye solution is circulated using dye circulator (Radiant Dyes RD 1000) and the output wavelength is characterized by a compact CCD spectrometer (AvaSpec-HS2048XL-EVO). The optical power of the pump laser and the dye laser is measured using a thermal sensor (Thorlabs PM100D). {\color{black}The output power of the latter is determined with a relative accuracy better than {1 \%}, which is required to normalize data taken at different excitation wavelengths.} Photoluminescence spectroscopy on the SWCNT sample is performed using a liquid nitrogen cooled photodiode (Horiba DSS IGA010L) attached to commercial spectrograph (Horiba Jobin Yvon IHR320). We also characterized the time resolution of the pump pulse and the dye-laser pulse with fast photodiodes (Thorlabs DET210 and DET410, respectively).

\paragraph{Optical configuration of the dye laser}

\begin{figure}[h!]
	\includegraphics*[width=\linewidth]{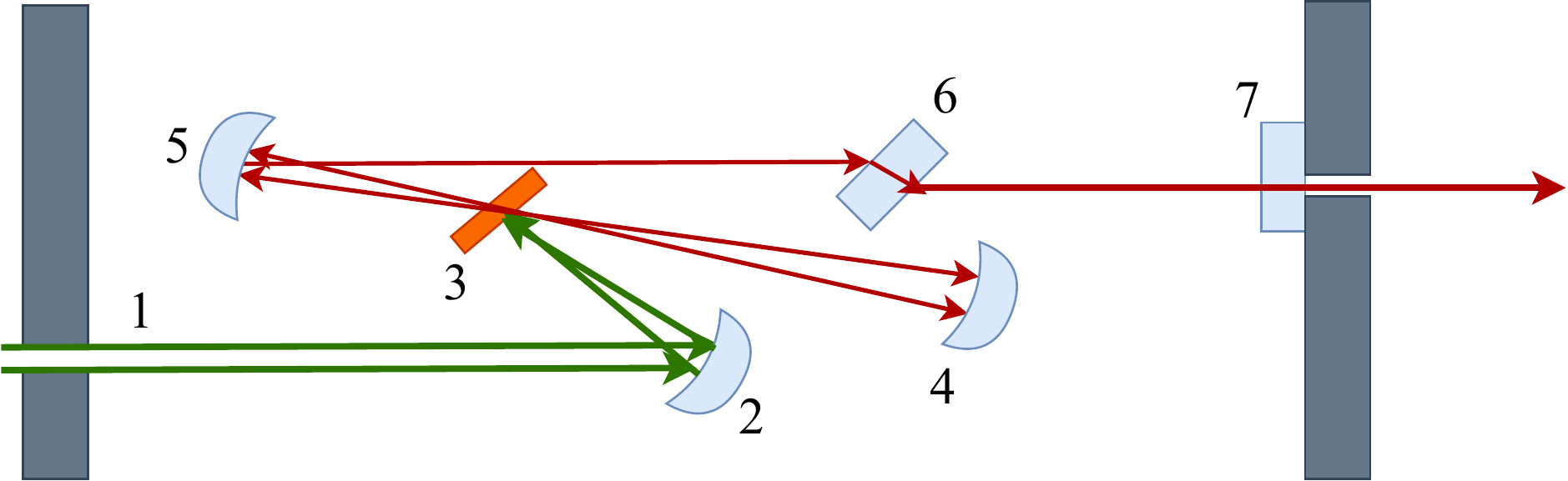} 
	\caption{Schematic diagram of dye laser resonator. 1: input aperture, 2: pump mirror ($f=50$ mm), 3: dye jet, 4: lower folding-mirror ($f=50$ mm), 5: upper-folding mirror ($f=50$ mm), 6: Lyot filter, 7: planar outcoupling mirror (Reflectivity $95\%$).}
	\label{rezo}
\end{figure}

Fig. \ref{rezo}. shows the configuration of the tunable laser resonator. The pump laser beam goes through the input aperture (1) and is focused on the dye jet (3) by the concave pump mirror (2), whose focus distance can be adjusted by a micrometer screw. The dye jet is a planar rectangular surface (with a Brewster angle with respect to the dye laser beam) with a thickness of approximately $200\ \text{$\mu$m}$, formed through a nozzle by the high pressure circulator. The active laser volume has a linear dimension of about $10\,\mu\text{m}$. The high reflectivity ($>99.8\%$) lower- (4) and upper-folding mirrors (5) collect the fluorescent light emitted by the dye. The lower folding-mirror refocuses the collected light to the jet. For this purpose, the mirror is placed at twice its focal length.  After the jet, the upper-folding mirror makes the beam parallel. 

The Lyot filter (6) serves as a tuner in the cavity. It is made of quartz plates, which is known to be birefringent i.e. its wavelength dependent transmission depends strongly on the orientation of the filter plates with respect to the laser polarization. Our system contains single, double and triple Lyot filters which improve the laser line purity. A single Lyot filter is a thin quartz plate with a thickness of unity (1), a double Lyot filter has two quartz plates with 4:1 thicknesses and the triple Lyot filter has three quartz plates with thicknesses of 16:4:1. We found that even a single Lyot filter results in about 2 cm$^{-1}$ FWHM (full width at half maximum) of the laser line, which is reduced below 1 cm$^{-1}$ with the triple Lyot filter. Given the about {\color{black}20-30 nm} spectral width of the SWCNT photoluminescence lines, this is more than sufficient spectral purity for the targeted purposes. In fact, this spectral purity would allow us to perform high resolution Raman spectroscopy on the SWCNTs.

The resulting laser beam leaves the resonator through the outcoupling mirror (7), which has a transmission of about 5\% {\color{black}for the 500-680 nm wavelength range}. In practice, one needs to use a set of "green-optimized" {\color{black}(Reflection $>99.99\,\%$ for 500-670 nm) and "red-optimized" (Reflection $>99.99\,\%$ for 570-740 nm)} lower- and upper-folding mirrors depending on the working wavelength, due to the limited wavelength coverage of high reflection laser mirrors. The dye laser system is used in combination with a home-built photoluminescence spectrometer which is described in Ref. \onlinecite{NegyediRSI}.

\paragraph{Dye preparation}
We used ethanol to dissolve Rhodamin 110 (also known as Rh. 560, 2-(6-amino-3-imino-3H-xanthen-9-yl)-benzoic acid), Rhodamin 6G (Rh. 590, 2-[6-(ethylamino)-3-(ethylimino)-2,7-dimethyl-3H-xanthen-9-yl]-benzoic acid) and DCM (4-(Dicyanometh-ylene)-2-methyl-6-(4-dimethylaminostyryl)-4H-pyran), purchased from Radiant Dyes GmbH, Germany. We found that the optimal solvent for Pyridin 1 (LDS 698, 1-Ethyl-2-(4-(p-dimethylaminophenyl)-1,3-butadienyl)-pyridinium
perchlorate) is benzyl-alcohol. In all cases, $500\ \text{mg}$ of the dye powder is dissolved in $50\ \text{ml}$ of solvent using a bath-sonicator for $15$ minutes. We fill the dye circulator with ethylene glycol (approx. $400\ \text{ml}$) and add the dye solution slowly, until we reach maximum output power of dye laser. To cover broad wavelength range one needs to use several dyes. To change the dye, the circulator is emptied and washed repeatedly by methanol and by circulating pure solvent. The cleaning {\color{black}of the circulator} is crucial especially when trying to use a "greener" dye after using a "red" dye: the green dye emission spectrum can overlap with the absorption spectrum of the red dye thus the latter can absorb the photons emitted from the green dye and it lowers the overall yield of the dye laser system.

\paragraph{SWCNT sample preparation}
We used single-walled carbon nanotubes prepared by the HiPco process (manufactured by the Carbon Nanotechnologies Inc.) with an average diameter of $1\ \text{nm}$ and $0.1\ \text{nm}$ variance. The abundance of SWCNTs with a different diameter (and hence different $(n,m)$ indices) can be described by Gaussian distribution. This SWCNT powder was mixed with $2\,\text{wt}\%$ solution of sodium deoxycholate (DOC) in distilled water. The approximate concentration of SWCNT in this solution was $1\,\text{mg/l}$. After 5 hours of tip-sonication (Branson SLPe) we placed the suspension into an ultracentrifuge (Thermo Scientific Sorvall MX 120) at $400\ \text{kg}$ for half an hour. The resulting decanted suspension contains individual nanotubes covered by the surfactant. The samples were then placed in quartz capillaries and sealed with a torch while the sample was kept in liquid nitrogen. We found that such sealed samples can be stored for an indefinite duration of time.

\section{Results}
\subsection{Performance and stability of the pulsed dye laser}

\begin{figure}[h!]
	\includegraphics*[width=\linewidth]{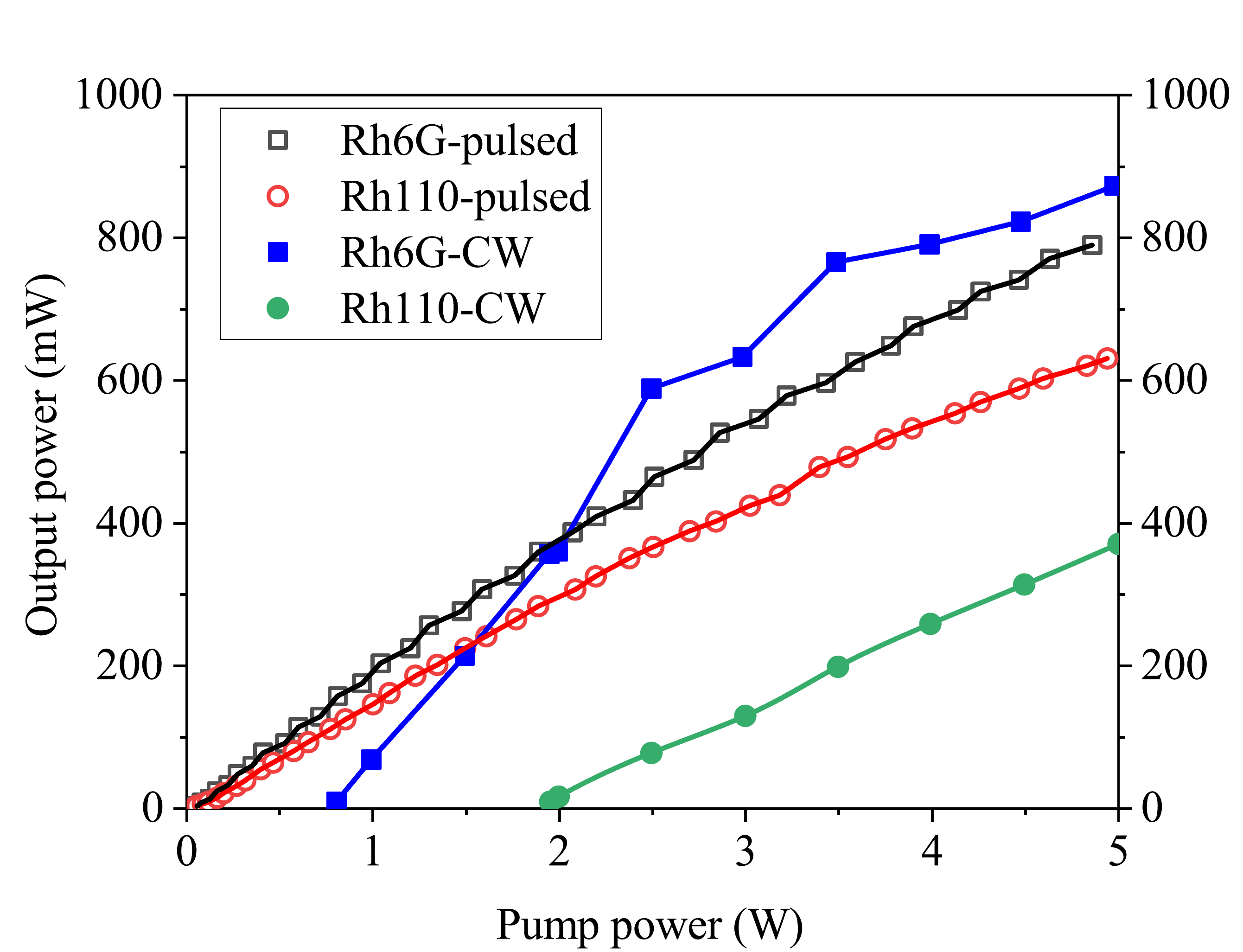}
	\caption{Comparison of dye laser pumped with CW laser and pulsed laser. In case of CW pumping a significant threshold is present. The repetition rate of the pulsed laser was set to $10\ \text{kHz}$, which is essential, as discussed below.}
	\label{threshold}
\end{figure}

The most important challenge, when using dye lasers, is to achieve operation for the less effective dyes. It is well known that e.g. Rh6G is an effective dye (due to a high, almost 100\% quantum efficiency) whereas e.g. Rh110, Pyridin 1 and 2 have lower performance. In practice, setting up a dye laser with Pyridin 1 is difficult. An important property of any pumped laser system is that lasing occurs only above a sizable threshold due to self absorption of the emitted photons \cite{duarte1990dye}. 

In Fig. \ref{threshold}, we compare the output power of the dye laser system pumped with the CW and the pulsed laser. The data is shown as a function of the exciting power, i.e. for the pulsed laser it corresponds to a large peak power. For the CW pumped operation, the threshold can be clearly observed, which is followed by a more or less linear dependence of the output power as a function of the pump power. The lower threshold value of Rh6G {\color{black}is related to the fact} that it is a more efficient dye with higher quantum yield than Rh110.

However, the threshold is practically zero for the pulsed laser pumping. Considering a $10\ \text{kHz}$ repetition frequency and 100 ns long pulses, we obtain that the peak power of the pulsed pump laser is 1000 times larger than that of the CW pump laser at the same average power. It means that the threshold is expected to be 1000 times smaller for the pulsed than for the CW pump laser. This is compatible with our experimental data as we observe that lasing of the dye laser starts already around 10 mW, which we consider as being essentially threshold-free operation. {\color{black}The fact that our laser operates a few 1000 times above the threshold power also affects the stability of the operation: it is known that transversal laser modes and also photons amplified after spontaneous emission compete with the desired laser mode \cite{LaserThreshold}.} 

{\color{black}We found that for the CW pumping, the noise of the tunable laser output is dominated by the latter effect, especially around the operating wavelength limits. The observed power noise is then much larger than the power noise of the pump laser (1 \% peak-to-peak power noise as specified by the manufacturer). The Q switch pump laser has a similar power noise (pulse-to-pulse RMS energy stability is better than 0.4 \%) and the tunable laser output noise essentially follows this limit. The additional noise due to spontaneous emission appears to be absent.}

The CW and pulsed pump lasers have fairly different beam characteristics, which is described by the so-called M$^2$ parameter (or beam propagation ratio). It is the product of the beam radius, $r$ and the half angle of divergence, $\theta$ as: M$^2=\frac{\theta r \pi}{\lambda}$. It essentially gives to what extent the beam deviates from a perfect Gaussian beam, which could be focused to a diffraction limited Airy spot. The CW laser has M$^2 \lesssim 1.1$, whereas the pulsed laser has M$^2\approx 20$. We note that this large M$^2$ prevented us from using a CW Ti:Sapphire laser in pulsed mode similarly to the dye laser system due to the too large irradiated volume inside the Ti:Sapphire crystal which was not compatible with the construction of the laser. For the dye laser system, the larger M$^2$ results in an outcoming beam with a somewhat inferior beam quality as compared to pumping with the CW laser. The usual CW pumped dye laser has an outcoming beam of about 2.25 mm diameter, which does not have an observable divergence after 1 meter (this is the practical distance of our laser usage conditions). In contrast, the output beam diameter of the pulsed dye laser is about 6 mm, which diverges to about an 8 mm diameter size after 1 meter. This is still a usable laser beam size for our spectrometer operation \cite{NegyediRSI}.

\subsection{Role of the repetition frequency}

\begin{figure}[h!]
	\includegraphics*[width=\linewidth]{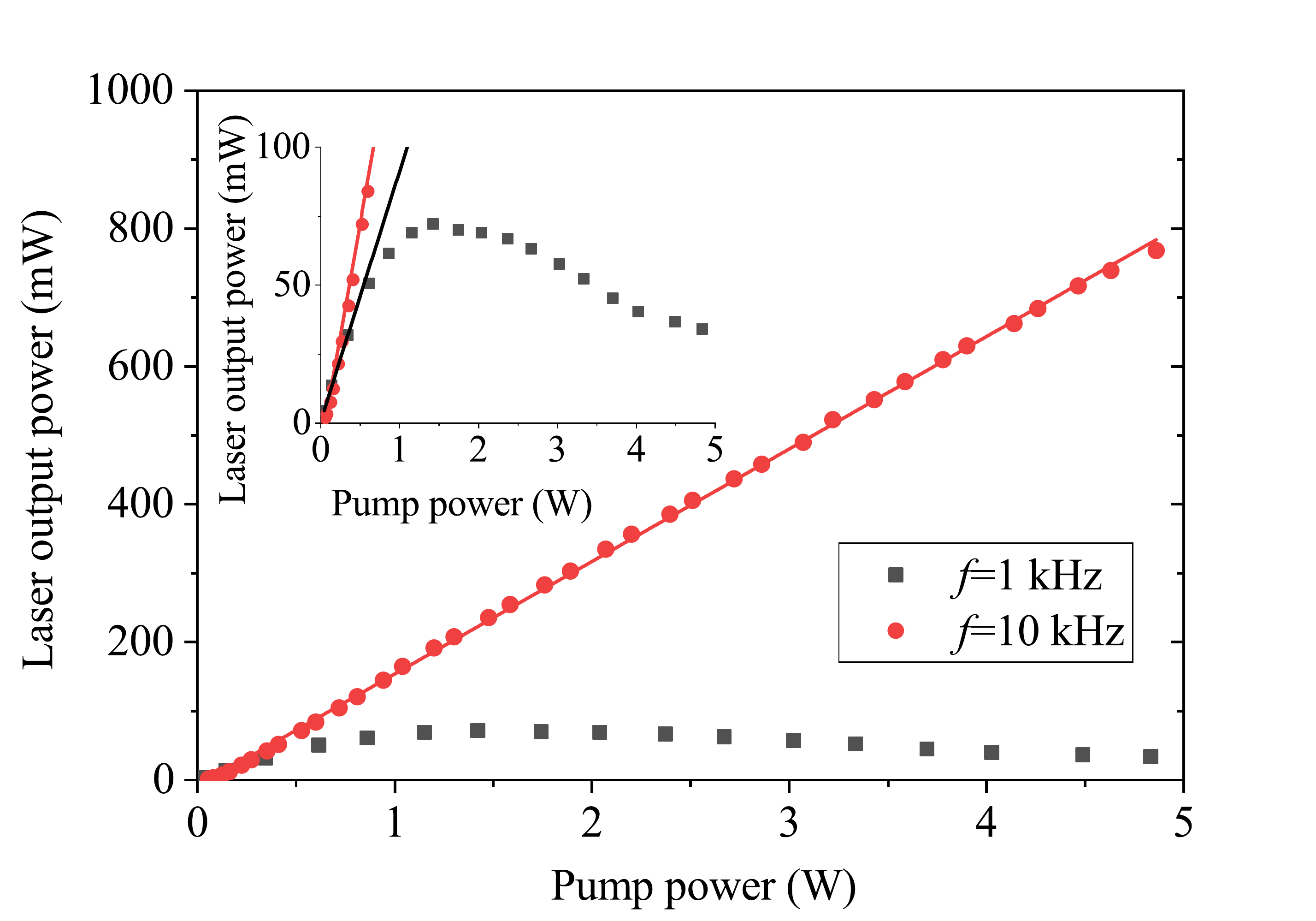}
	\caption{Output power of pulsed dye laser with $1\ \text{kHz}$ and $10\ \text{kHz}$ repetition rates. The inset shows a close up of the curves at low powers. Note that the slope of output power is identical for both repetition rates at low powers but the dye laser output rapidly saturates above a pump power of 1 W for the 1 kHz repetition. }
	\label{pow_1_10khz}
\end{figure}

The pump laser can be operated with three different pulse repetition rates: $1\ \text{kHz}$, $5\ \text{kHz}$ and $10\ \text{kHz}$. Fig. \ref{pow_1_10khz} shows the measured dye laser output power as a function of pump power for the $1\ \text{kHz}$ and $10\ \text{kHz}$ repetition rates, while the settings of the optical resonator remained unchanged.
At the $1\ \text{kHz}$ repetition rate, the output power rapidly saturates for a pump power above 1 W, as the inset of Fig. \ref{pow_1_10khz} shows. It is caused by the high energy of the incoming pulses which probably either optically saturates the active medium of the dye laser or even damages it. The laser dyes are also known to become opaque for high pumping powers due to the excitation of a higher lying triplet level \cite{duarte1990dye}. The excited state of the dyes can cross over to a triplet state, whose optical absorption matches the emitted light range, which thus prevent lasing. The stronger the optical pumping is, the larger the probability of the triplet crossover.

In contrast, the saturation effect is absent for the $10\ \text{kHz}$ repetition frequency at least in the pump power range which we studied. We observed that the output power follows linearly the pump power therefore this repetition frequency is optimal for our system.

\subsection{Tuning range of the pulsed dye laser}

\begin{figure}[h!]
\includegraphics*[width=\linewidth]{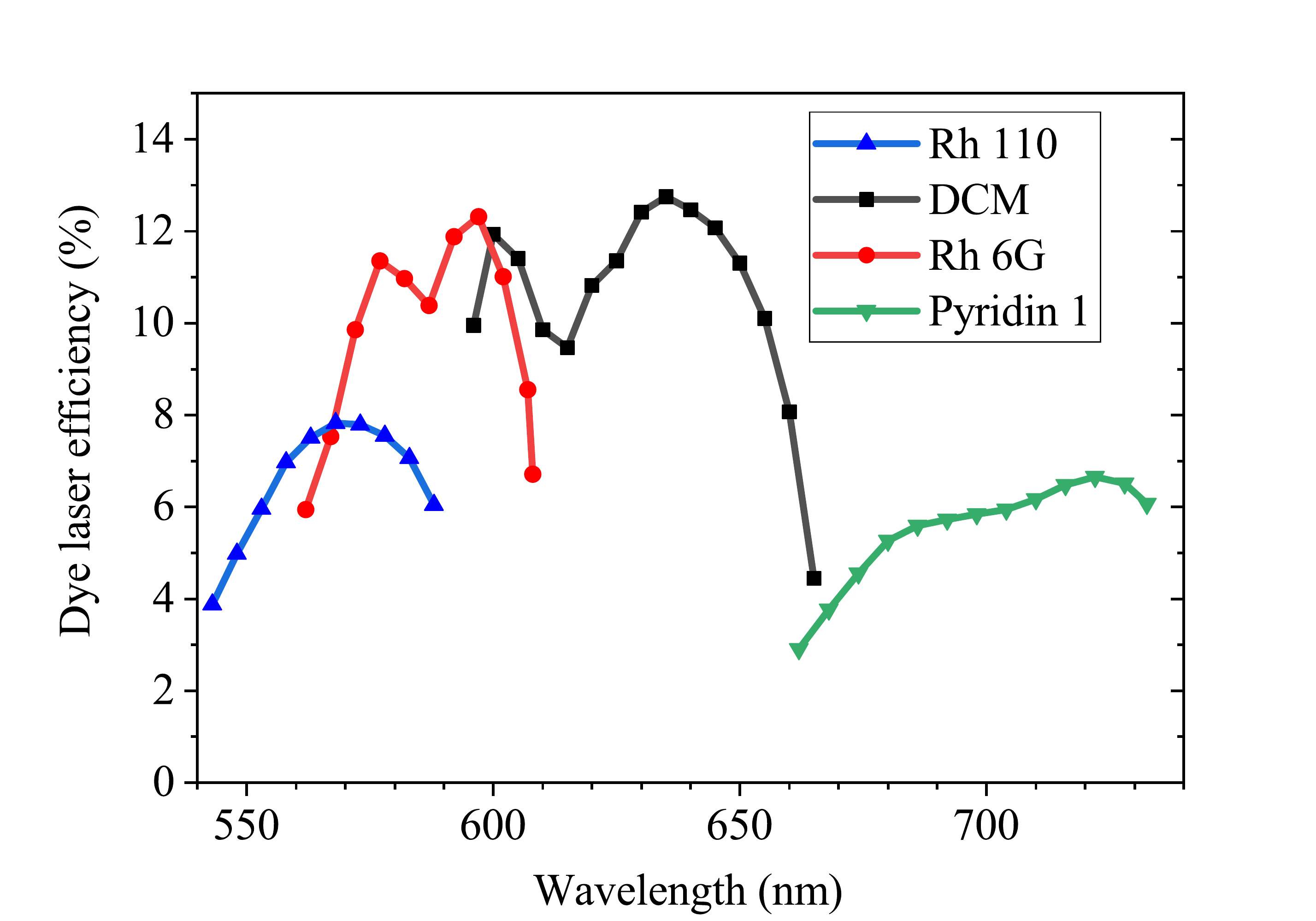}
\caption{Dye laser efficiency for the various dyes as a function of the emitted wavelength. Note that the four dyes allow to cover a wavelength range of about $200\ \text{nm}$.}
\label{tune}
\end{figure}

Fig. \ref{tune} shows the dye laser efficiency (i.e. dye laser output power divided by the pump power) measured in $6\ \text{nm}$ steps as a function of wavelength. The wavelength range is limited by the luminescence spectrum of the given dye and the wavelength dependent gain of the dye laser resonator. Under CW excitation, our system works well with the Rh6G and DCM dyes. However operation with the Rh110 and Pyridin 1 has always turned out to be difficult and required a substantial effort. With the present improvement, the pulsed excitation at a repetition frequency of 10 kHz allowed for a relatively effortless operation with these two dyes, too. We note that the use of the four dyes allows to cover the wavelength range essentially from the exciting wavelength up to the range where Ti:Sapphire lasers operate well.

We underline that pulsed dye lasers have long been known \cite{Haensch1972} and in fact their operation is often more convenient than that of CW lasers. The most important result of our work is that we improve the stability of a CW laser for several dyes for which the operation is otherwise difficult. From a technical point of view, we still operate the pulsed dye laser as if it was a CW laser as we integrate the resulting PL signal.

\subsection{Characterization of laser pulses} \label{char}
\begin{figure}[h!]
	\includegraphics*[width=\linewidth]{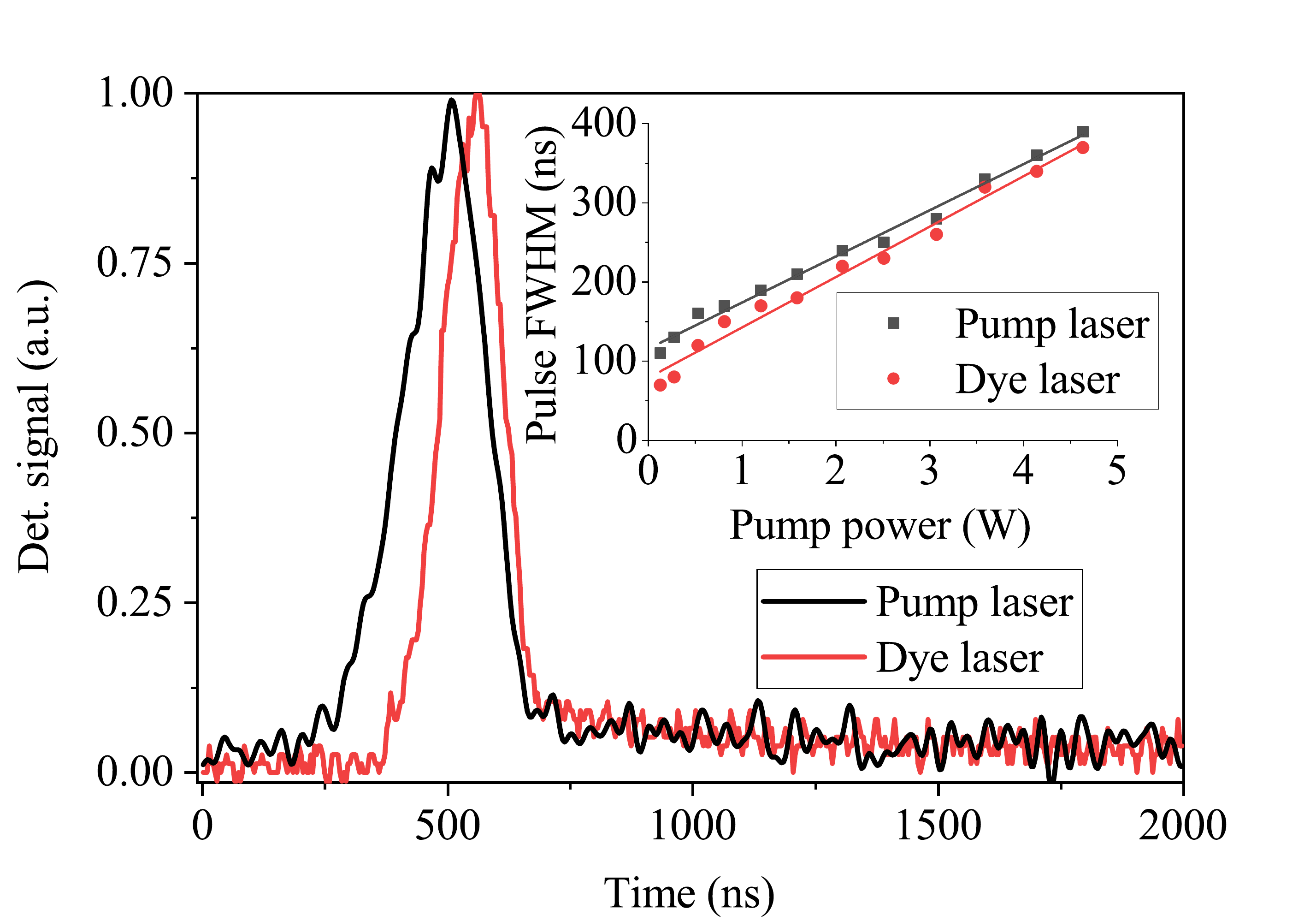}
	\caption{Pump pulse and dye laser pulse measured using fast photodiodes. The dye laser pulse follows the pumping pulse and has the same shape in time domain. For clarity, the two signals are scaled to have the same amplitude. The inset shows pulse length as function of pump laser power measured on pump pulse and dye (DCM) laser pulse at the same time.}
	\label{pulses}
\end{figure}

In Fig. \ref{pulses}, time resolved detection of laser pulses is presented. A thin quartz plate acts as a beam sampler in the path of laser light to outcouple a small amount of laser light to the photodiode. The dye laser pulse follows the pump pulse with a delay of $40\ \text{ns}$. We checked carefully for the possible origin of this short time lag in the light detection electronics (by using matched cable lengths for the two detectors and by properly impedance matching the photodiodes) but it did not reveal any artefacts. We therefore conclude that the time delay between the pump and dye laser output originates from the laser operation itself. 

The inset of Fig. \ref{pulses} shows the measured pulse lengths as a function of the pump power. Increasing the power broadens the pump pulse and hence the width of the dye laser pulse. The dye laser pulse always remains shorter than the pump pulse. {\color{black}This effect is probably related to the lifetime of the excited state in dye molecules, which is not longer than a few nanoseconds \cite{duarte1990dye}.}

The time resolution of our setup is limited by the pump pulse length. Therefore time resolved measurements on carbon nanotubes did not reveal any information on the dynamics of luminescence, which are known to have much shorter luminescence lifetimes \cite{ReichBook}. The repetition rate gives an upper limit for excitation lifetime of the studied system because it has to relax between two pulses. In our case, this means that the system is suitable for time resolved measurements in the range from $1\,\mu\text{s}$ to $1\ \text{ms}$

\subsection{Photoluminescence measurements on SWCNTs with the pulsed dye laser}

\begin{figure}[h!]
	\includegraphics*[width=\linewidth]{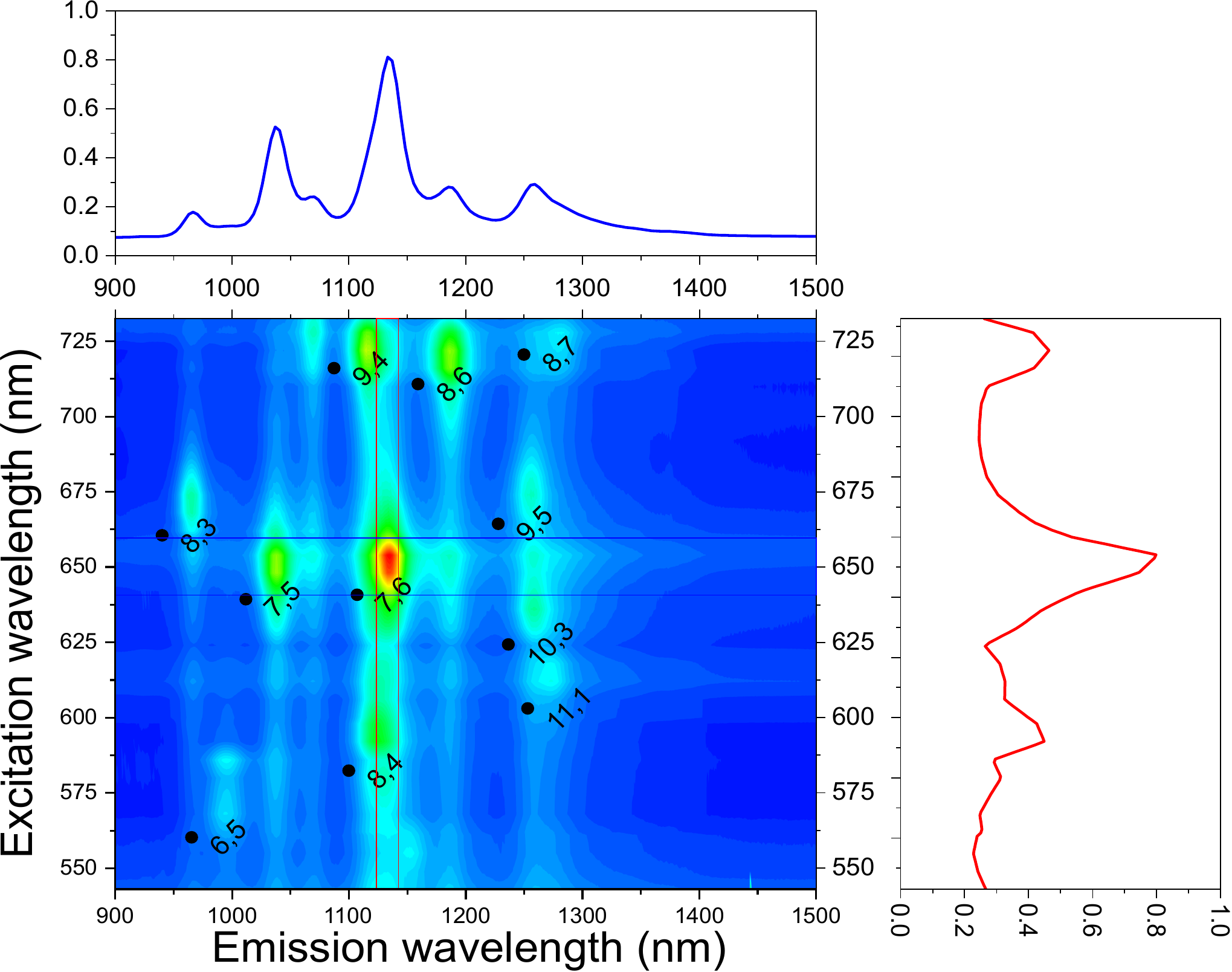}
	\caption{Photoluminescence map of an SWCNT sample measured at room temperature. Individual peaks can be well observed due to the great contrast and spectral resolution of the spectrometer. The data were taken with 6 nm excitation wavelength resolution. The $n,m$ SWCNT indices are given from Ref. \onlinecite{WeismanNL2003}, the discrepancy is due to the different surfactant used. A vertical and a horizontal profile is also shown.}
	\label{pulsed_map}
\end{figure}

In Section \ref{char}., we presented that a broad wavelength range can be covered with our pulsed dye laser system. We underline that the original CW laser system could only be operated down to about 560 nm and up to 670 nm, however the present improvement extends it to a 540-730 nm operation. In Fig. \ref{pulsed_map}., the individual photoluminescence spectra measured at different excitation wavelength with a 6 nm resolution are combined to obtain the so-called photoluminescence map \cite{Bachilo:Science298:2361:(2002)}. Acquiring such maps is time consuming with the present setup and is presented for demonstration purposes.
{\color{black} We used typical incident laser powers of 10-50 mW which is focused to a spot diameter of about $20\,\mu\text{m}$. We checked that this power density on the sample does not cause any non-linear effects even under the pulsed operation conditions.} {\color{black}However, care must be taken for such a high temporal laser power density as it could lead to e.g. trion generation \cite{Santos}.}

{\color{black}We note that the so-called supercontinuum lasers \cite{supercontinuum} emerge as an alternative for conventional tunable lasers (such as used herein) for photoluminescence mapping \cite{SuperCont_PL1,SuperCont_PL2}. 
However, supercontinuum lasers have a smaller power spectral density (about 1-5 mW/nm), whereas in our case its CW averaged value is about 100-200 mW/nm (we normalized our power for the effective observed spectral window, the laser line itself is as sharp as 0.1 nm).}

The major limitation is the time taken for the dye exchange and some inevitable fine tuning of the laser. We believe that our system may find application in cases where subtle details of the map are to be studied, where the incoming laser brilliance can be a decisive factor, e.g. when studying weakly allowed electronic transitions or vibrational modes \cite{DresselhausPRL2005}, electron-phonon interactions \cite{PhysRevLett.94.127402}, or other weak modes e.g. after chirality selective separation \cite{KatauraSortingNatComm2011,KatauraSortingNatComm2016}. 

\section{Summary}

We presented the development and the performance characterization of a pulsed dye laser based photoluminescence spectrometer. We investigated the power output of the pulsed dye laser and compared it to the same system, when pumped by a continuous wave laser. We found that the laser operation is significantly simplified and is enabled even for dyes which are {\color{black}otherwise on the border of operability of our CW laser.} This is due to the high peak power of the pumping pulses which allows a threshold free operation. The optimum peak power can be conveniently controlled for a given average power by varying the pulse repetition rate of the Q-switch pump laser, for which we established the optimum conditions. The use of four dyes allows to cover the 540-730 nm wavelength range. Our time resolved detection of the dye laser output shows that the dye laser pulse follows the pump pulse. We demonstrated the tuning ability of our light source by presenting a photoluminescence map on an SWCNT sample.

\section{Acknowledgement}
Work supported by the Hungarian National Research, Development and Innovation Office (NKFIH) Grant Nr. K119442 and 2017-1.2.1-NKP-2017-00001. The research reported in this paper was supported by the BME-Nanonotechnology FIKP grant of EMMI (BME FIKP-NAT).

\bibliography{pulsed,Tubes2012June}

\end{document}